# A Novel Clustering Algorithm for Coverage a large scale in Wireless Sensor Networks

Wassim JERBI[1], Abderrahmen GUERMAZI[2] and Hafedh TRABELSI[3]

[1,2]Higher Institute of Technological Studies, 3099 El Bustan Sfax, Tunisia.

and

[3]CES research unit, National School of Engineering of Sfax, University of Sfax, Tunisia.

## ABSTRACT

*The applications require coverage of the whole monitored area for long periods of time. Clustering is a way to reduce communications, minimize energy consumption and organize messages among the cluster head and their members. The message exchange of communication and data transmission between the different sensor nodes must be minimized to keep and extended the lifetime of the network because of limited energy resources of the sensors. In this paper, we take into consideration the problem isolated nodes that are away from the cluster head (CH) and by consequence or CH is not within the reach from these nodes. To solve this problem, we propose O-LEACH (Orphan Low Energy Adaptive Clustering Hierarchy) a routing protocol that takes into account the orphan nodes. Indeed, a cluster member will be able to play the role of a gateway which allows the joining of orphan nodes. Our contribution is to election a cluster head that has enough energy for a better now to coordinate with these member nodes and maintain the full coverage for applications which requires of useful data for the entire area to be covered. The simulation results show that O-LEACH performs better than LEACH in terms of connectivity rate, energy, scalability and coverage.*

## 1. INTRODUCTION

The technological advances in microelectronics have allowed the birth of the small sensor capable of handling communications without wireless with a limited energy. Clustering is a set of sensors, linked together by a cluster head, to form a group. A sensor will randomly elected a cluster head. After the selection of cluster head, each sensor can be communicated locally by the leader of the cluster. LEACH [1] is considered as the basic hierarchical routing protocol (cluster-based approach). It is also one of the most popular cluster based routing algorithms for (Wireless Sensor Networks) WSNs. It combines both the efficiency in energy consumption and the quality of access to the media, and it is based on the division into groups, with a view allowing the use of the concept of data aggregation for a better performance in terms of lifetime.

Cluster Heads (CH) are randomly chosen in a specific election algorithm based on a probability function that takes into account various criteria such as the available energy. The routing protocols are actually divided into two families: central data and hierarchical routing protocols. In a hierarchical topology, can be cited references protocols, Younis et al. (2004) have proposed a HEED [2], Lindsey et al. (2002) have proposed a PEGASIS [3], Manjeshwar et al.(2001) proposed a TEEN [4], and A Manjeshwar et al. (2001) proposed a PTEEN [5].

Leach performs the single-hop inter-cluster, directly from CHs to the BS, routing method, which is not applicable to large-region networks Akyildiz et al. (2002) [6]. It is not always a realistic assumption for single-hop inter-cluster routing with long communication range Al-Karaki et al (2004) [7]. Besides, long-range communications directly from CHs to the BS can breed too much energy consumption; despite the fact that CHs rotation is performed at each round to achieve load balancing, LEACH cannot ensure real load balancing in the case of sensor nodes with different amounts of initial energy, because CHs are elected in terms of probabilities without energy considerations Xuxun Liu (2012) [8]. The coverage preservation can achieve a better quality of service necessary to process communications between processes, base station, cluster head and sensor nodes for wireless sensor network. For this, the selection of adequate cluster head allows full coverage of the controlled area. But, this is not always the case, because we do not know how to select the cluster head, which leads to better structural organization of its cluster S. Soro and W. Heinzelman (2009) [9].

The idea of dynamic clustering brings extra overhead. For instance, CH changes and advertisements may diminish the gain in energy consumption Li, C et al. (2011) [9]. LEACH is very favorable in terms of energy efficiency. however, controlling the number and the location of the clusters head (CHs) and also the size of the clusters about the node number leads to a balance in energy use of the CHs and increasing the lifetime of the network, Asgarali Bouyer et al (2015)[10].

Nevertheless, in a round, the nodes which are not CH may not join a cluster. In such a case, the data which must be collected from the node outside the network (orphan node) could have a great importance in some applications. Hence, these applications will be concrete ones and will satisfy our needs. Obviously, we need to collect data from all distributed nodes inside the network, hence allowing taking the suitable decisions.

The large-scale deployment of controlled high Wireless Sensor Networks (WSNs) necessitates an efficient organization of the networks for high network connectivity and a low orphan node ratio. Where sensor network are randomly deployed, they are not uniformly distributed inside field. As a result, some places in the field don't benefit from a good connectivity. Hence, the routing protocols conceived for the WSN must have a self organization capacity in order to adapt them to the random distribution of the nodes and the dynamic topology of the network.

An orphan node which does not belong to any CH sends a message towards its nearest neighbors which belong to a cluster (belonging application). A member of the cluster will represent a gateway allowing the link between one or several orphan nodes and the CH.

Among the factors, we must verify that in each round, the number of distributed nodes is approximately equal to the actual number of connected nodes. If the number of connected nodes is less than the required number, the nodes that are not within the reach of CH are called orphan nodes.

In this paper, we propose a protocol called O-LEACH which allows joining the orphan nodes. To solve the problem, one node member of a cluster receives "Orphan notification". The member of the CH will be a gateway. The cluster member receives a request message from a node that belongs to any group and asks for a membership in this group. Different messages are transmitted between the three processes that are: the CH, the member nodes of the cluster (gateway) and the nodes without connectivity orphan nodes (orphan nodes). These transactions generate an adequate link between the orphan nodes and the cluster.

The remainder of this paper is organized as follows: section 2 describes the related work of routing in WSN and emphasizes on existing CH selection method. Section 3 describes O-LEACH protocol. In Section 4, we present the performance evaluation of O-LEACH and its comparison with LEACH. Section 5 concludes the paper.

## 2. RELATED WORK ON ROUTING PROTOCOL

### 2.1. Protocol LEACH

In WSNs, the use of routing protocols designed for the traditional ad hoc networks is inappropriate. This is due to the characteristics allowing distinguishing the two types of networks. Hence, we need to improve or develop new specific routing protocols for WSN. LEACH is considered as the first hierarchical routing protocol. It is also one of the most popular hierarchical routing algorithms for WSN, proposed as part of the project. It combines the efficiency in energy consumption and the quality of access to the media, and it is based on the division into groups, in order to allow the use of the concept of data aggregation for a better performance in terms of lifetime.

Heinzelman proposed that the LEACH cluster formation is made by a centralized algorithm at the base station (BS).
The aim of the LEACH protocol is to form clusters based on the intensity of the received radio signal. Indeed, LEACH uses an algorithm which is distributed where each node decides autonomously whether it will be a Cluster head or not by randomly calculating a probability pu and comparing it to a threshold T(u); Then, it informs its neighborhood about its decision. Each node decides which Cluster head to join by using a minimum transmission of energy (i.e the nearest). The algorithm consists of several rounds and, for each round, a rotation of the role of the Cluster head is initiated according to the probability pu which is chosen and compared to the following formula of the threshold:

$$T(n) = \frac{P}{1 - P * \left( r \bmod \frac{1}{p} \right)} \; if \; n \in G, \qquad (1)$$

$$T(n) = 0 \qquad \text{otherwise}$$

p: the percentage of CHs on the network, r : the current round number and G: the set of nodes that was not CH in the (1/p) preceding rounds.

During a period T, a node n chooses a random number of x whose value is between 0 and 1 (0<x <1). If x is less than a threshold value, then the node n will become a cluster head in the current period. Otherwise, the node n should join the nearest cluster head in its vicinity.

Rounds in LEACH have predetermined duration, and have a set-up phase and a steady-state phase. Through synchronized clocks, the nodes know when each round starts.

We use for our consumption model energy the same first order radio model is represented in [1]. Transmitting L bit in the packet data is described as follows:

$$ETx(l, d) = ETx\text{-}elec(l) + ETx\text{-}amp(l, d) \qquad (2)$$

$$= \begin{cases} lEelec + l\varepsilon fsd2 & d < d0 \\ lEelec + l\varepsilon ampd4 & d \geq d0 \end{cases} \qquad (3)$$

The power consumption of receive packet L-bit is:

$$(4)$$

$$ERx(l) = ERx\text{-}elec(l) = lEelec$$

d is the propagation distance,

$$d_0 = \sqrt{\varepsilon_{fs}/\varepsilon_{amp}} \qquad (5)$$

And the symbol Eelec represents the energy consumption of radio dissipation, εfs and εamp represent the energy consumption of amplifying radio.

## 2.2 Set-up Phase

This phase starts by the announcement of the new round by the sink node, and by taking the decision for a node to become a CH with a probability pi(t) in the beginning of round r+1 which starts at the instant t. once a node is chosen CH, it must inform the other CH nodes about its position in the current round. For this, a warning message ADV containing the identification of the CH is transmitted to all CH nodes by using the protocol MAC CSMA () in order to avoid the collisions between the various CH. Each member informs its CH about its decision.

After the grouping operation, each CH acts as a local control centre in order to ensure the coordination between the data transmission inside its group. It creates a schedule TDMA and assigns to each member node a slot for data transmission. The set of slots assigned to the group nodes is called a frame.

Figure 1 shows the formation of clusters in a round. Each cluster includes a CH and some pickup member nodes.

## 2.3 Steady State Phase

This phase is longer than the preceding one, and allows the collection of the received data. By using the TDMA, the members transmit their received data during their slots. This allows them to switch off their communication interfaces outside their slot in order to save energy. Then, these data are aggregated by the various CH which merge and compress then before sending the final result to the sink node.

After a predetermined time, the network will go through a new round. This process is repeated until the moment where all the nodes of the network will be chosen CH, one time, all throughout the preceding rounds. In this case, the round is again initialized to zero.

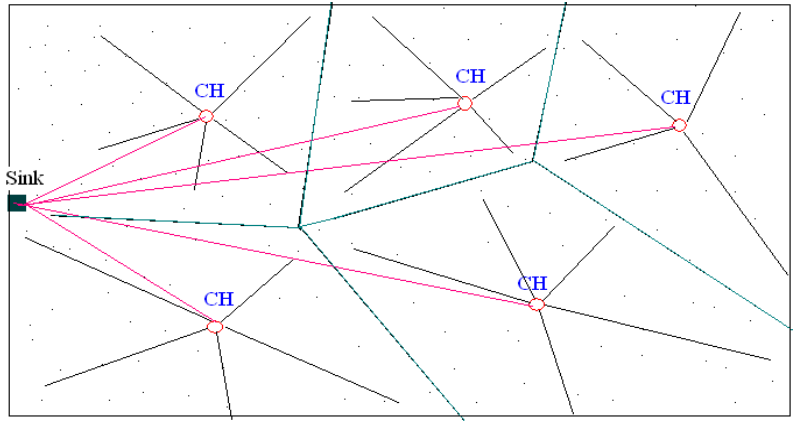

Figure 1.  T round in LEACH

## 2.4 Advantages of LEACH

Le protocole LEACH engendre beaucoup d'avantages en ce qu'il offre comme bonne manipulation de ressources du réseau en respectant plusieurs contraintes telle que la consommation d'énergie.

- Protocole auto-organisateur basé sur le groupement adaptatif: LEACH est complètement distribué, autrement dit, les nœuds prennent leurs décisions de façon autonome et agissent de manière locale et n'ont pas besoin d'une information globale ni d'un système de localisation pour opérer de façon efficace. De plus, la collection de données est faite périodiquement (l'utilisateur n'a pas besoin de toutes les données immédiatement). Pour exploiter cette caractéristique, ce protocole introduit un groupement adaptatif, c'est-à-dire, il réorganise les groupes après un intervalle de temps aléatoire, en utilisant des contraintes énergétiques afin d'avoir une dissipation d'énergie uniforme à travers tout le réseau.
- Rotation des rôles de chefs de groupes: La rotation des rôles de chefs de groupes s'avère un facteur important pour l'organisation des nœuds. Ce rôle est épuisant en termes de d'énergie car les CH sont actifs tout au long de leur élection. Puisque le nœud puits est généralement loin du champ de surveillance, les CH diffusent une quantité plus importante d'énergie pour lui transmettre leurs données. Donc, si les CH sont choisis d'une manière fixe, leur énergie s'épuisera rapidement ce qui induit à leur défaillance. Par conséquent, tous les autres nœuds seront sans CH et donc inutiles. C'est pourquoi, les algorithmes de groupement (*clustering*) étudiés jusqu'ici adoptent la rotation du rôle de chefs de groupes.
- Faible énergie pour l'accès au média: Le mécanisme de groupes permet aux nœuds d'effectuer des communications sur des petites distances avec leurs CH afin d'optimiser l'utilisation du média de communication en la faisant gérer localement par un CH pour minimiser les interférences et les collisions.

- Compression locale (agrégation) : Les CH compressent les données arrivant de leurs membres, et envoient un paquet d'agrégation au nœud puits afin de réduire la quantité d'informations qui doit lui être transmise. Cela permet de réduire la complexité des algorithmes de routage, de simplifier la gestion du réseau, d'optimiser les dépenses d'énergie et enfin de rendre le réseau plus évolutif (scalable).

## 2.5. Disadvantages of LEACH

In what follows, we present the advantages as well as disadvantages of the LEACH protocol.
During a round, we may not have any CH if the random numbers generated by all the nodes of the network are higher than the probability pi (t).

- The farthest nodes form the CH die rapidly as compared with the nearest ones.
- The use of a communication with one jump instead of a communication with several jumps reduces the nodes energy.
- The LEACH protocol cannot be used in the real time applications since it leads to a long period.
- The rotation of the CH ensures not to exhaust the batteries. However, this method is not efficient for networks with a big structure because of the overflow of announcements generated by the change of the CH, hence reducing the initial energy gain it's not obvious to have a uniform distribution of the CH.
- As a result, it is possible to have the CH concentrated in one part of the network. Hence, some nodes won't have any CH in their neighbourhood.
- LEACH is suitable for small size networks because it assumes that all nodes can communicate with each other and are able to reach sink, which is not always true for large size network.
- Since CH election is performed in terms of probabilities, it is hard for the predetermined CHs to be uniformly distributed throughout the network. Thereby there exist the elected CHs that are concentrated in one part of the network and some nodes that have not any CHs in their vicinity Seah (2010) [15].

## 3. DESCRIPTION OF O-LEACH PROTOCOL

### 3.1 Orphan Problem and Proposed Solution

In Wireless Sensor Networks, low latency, energy efficiency, and coverage problems are considered as three key issues in designing routing protocols Wafa Akkari et al (2015)[12]. The choice of the optimal routes ensures the delivery of information to the base station and reduces the packet delivery delay. Thus, the network must pass across by maximizing the networks life without decreasing its performance. The drawback of the LEACH protocol is the limited use in a wide field since many remote CH nodes die rapidly (as compared to a small field) because the nodes cannot join them.

The optimal percentage of the desired number of CH should be proportional to the total number of nodes. If this percentage is not met, this will lead to greater energy dissipation in the network. Indeed, if the number of CH is very high, there will be a large number of nodes (CH) dedicated to very expensive tasks in energy resources. Hence, there will be considerable energy dissipation in the network. Moreover, if the number of CH is very small, the latter will manage groups of large sizes. Thus, these CH will be consumed rapidly in case an important work is required from them. The routing protocols are actually divided into two families: central data and hierarchical routing protocols.

During the construction of the clusters the pickup nodes choose randomly the CH which can be concentrated in a specific part of the work field. As a result, the remaining part of the field will not be covered (see figure 2) the pickup nodes will be outside the network. The values received by the orphan nodes will not be transmitted to the base station. The problem of the orphan nodes requires finding a solution allowing to join these nodes to the remaining part of the network.

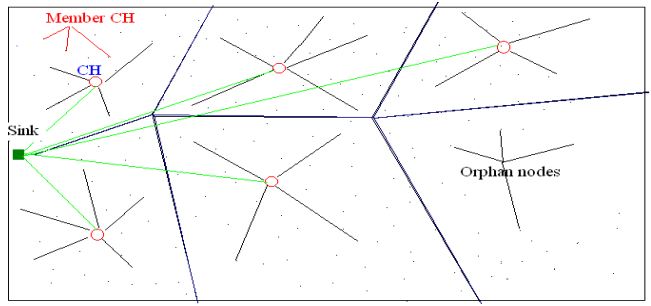

Figure 2. T round in O-LEACH

## 3.2 Set-up Phase Extension O-LEACH

The O-LEACH protocol consists of two phases: Set-up phase and steady-state phase as illustrated in figure 3 and figure 4:

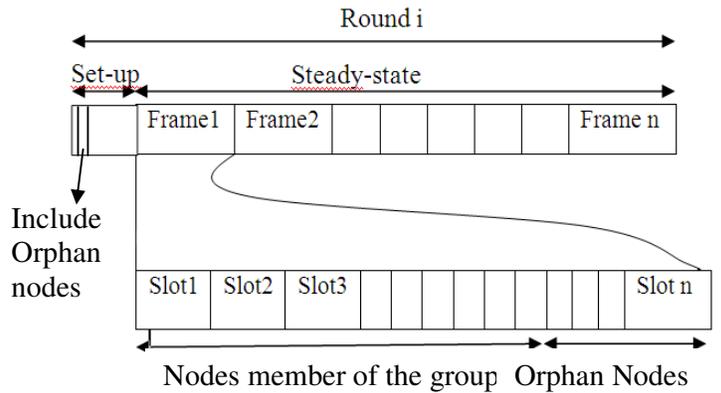

Figure 3. T round in O-LEACH reserved slots

The initialization phase consists in selecting the CH nodes with a certain probability, by the local decision taken by a node to become a CH. After the construction of clusters, a timer is used in order to verify the existence of orphan nodes.

If the answer is positive, CH gateway informs the CH' (the first orphan node having the access to the gateway) about the number of slots to be reserved to the orphan nodes by the CH. The CH' play the same role as the CH.

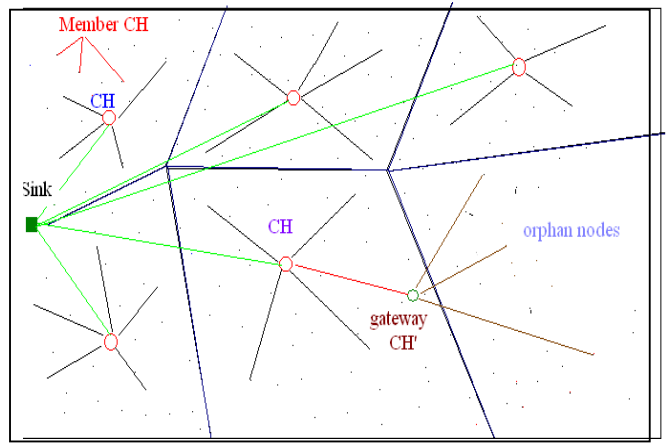

Figure 4. Solution with member gateway CH

## 3.3 Transaction message Proposed Algorithm

Three processes involved in the resolution of orphan nodes, which are orphan nodes, member cluster (Gateway CH') and CH node (figure 5):.

- Orphan nodes send status to member cluster.
- Member cluster says I am a Gateway.
- Orphan nodes join Gateway.
- The Gateway informs the CH node the number of slots.
- CH node reserve number of slots TDMA (Cluster + sub cluster).
- The gateway broadcasts a TDMA slot to each orphan node.

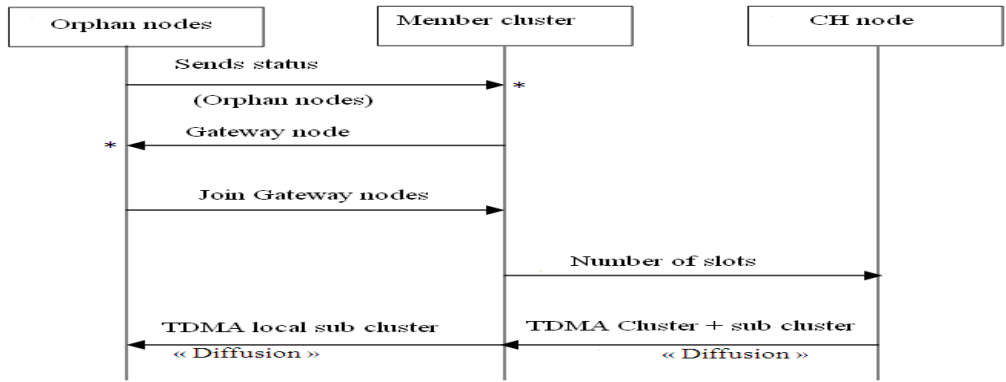

Figure 5. Transaction message

### 3.4 Flowchart of Proposed Algorithm

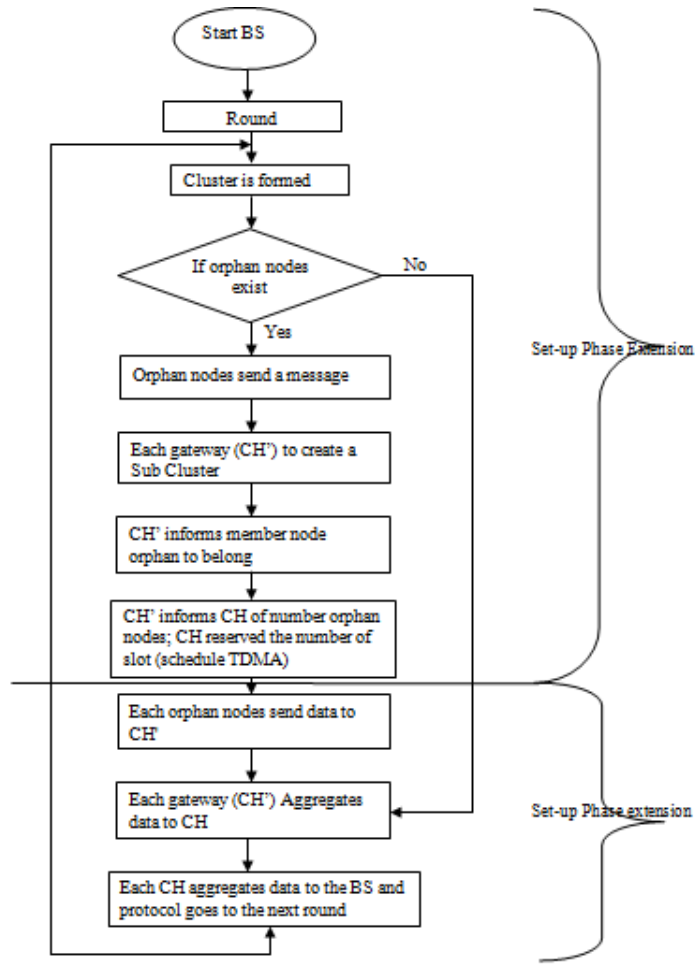

Figure 6. Flow chart of proposed algorithm

The proposed algorithm works in rounds. Each round performs these following steps, illustrated in Figure 6:

- Periodically, the base station starts a new round by incrementing the round number.
- The selection of the CH in the LEACH protocol according to a probability between 5%and 15%, the number of CH doesn't exceed the numbers 15 and 5 for each round (as related to the percentage used). For each round, a pickup node chooses a cluster as a head by selecting a random number to be compared to the threshold value. The threshold T(n) is set as: T(n) = {P / 1 – P * (r mod1/P)} if n belongs to G, if not zero. P is the desired percentage of cluster heads, r is the current round, and G is the set nodes that have not been cluster heads in the last (1/P) rounds.
- As soon as a cluster is formed, the number rounds wait for a message from nodes orphan.

- The node member of a cluster will be a gateway and will inform the orphan node to be a CH'.
- The construction of sub-clusters with a group leader CH'.
- The CH' gathers and aggregates the data toward the gateway.
- The gateway sends two slots to the CH, the first one includes the data of CH' whereas the second one contains the gateway data.
- The CH gathers the data, makes the necessary treatments and finally transmits them to the BS.

## 3.5 Steady State Phase O-LEACH

This phase allows the collection of the pickup data. By using the TDMA the member nodes of the cluster and the orphan nodes transmit their pickup data during their own slots. The node CH' gathers the pickup data from the neighbouring members. Then, these data are aggregated by the CH' which merge and compress them before sending the final result to the CH node through the gateway. The gathers the data of the pickup nodes (members of the cluster and orphan), aggregates and transmit then to the base station.

## 3.6 Pseudo Code

- Choosing Major Cluster Head and Choosing Cluster Head

```
{for each round  // round start
// Choosing Major Cluster Head
Number Major Cluster Head = 2% total of the nodes
{for Major Cluster Head
Major cluster head coordinate x
Major cluster head coordinate y
// Choosing Cluster Head and number of Cluster Head
   Threshold value is set to (P / (1 – P * (round % 1/P)))
   {for each node
// each node has a probability P of becoming a cluster-head
Rate of clustering (TR) =10%
{if number of cluster <= TR && energy of node > 0
Assign a random number of node
  {if (random number < threshold value) && (the node has not been cluster head)
    Node is Cluster head //assign node id to cluster head list Increment cluster   head
count //a new cluster head has been added
Else go to the next node  }
Else go to the next node  }
  }}
```

orphans nodes to join a gateway

{if distance between node and cluster head is <= the transmission range
Transmission cost is: d is the propagation distance,

$$d_0 = \sqrt{\varepsilon_{fs}/\varepsilon_{amp}}$$

Transmitting L bit in the packet data is described as follows:

$$E_{Tx}(l, d) = E_{Tx\text{-elec}}(l) + E_{Tx\text{-amp}}(l, d)$$

$$\begin{cases} lE_{elec} + l\varepsilon_{fs}\,d2 & d < d0 \\ lE_{elec} + l\varepsilon_{amp}\,d4 & d \geq d0 \end{cases}$$

The power consumption of receive packet L-bit is:

$$E_{Rx}(l) = E_{Rx\text{-elec}}(l) = lE_{elec}$$

Subtract the transmission cost from the sending nodes

{if remaining energy <= 0

   display node has dead
   exit}
   Subtract the reception cost from the receiving node

{if remaining energy <= 0

## 4. O-LEACH PERFORMANCE EVALUATIONS

### 4.1 Simulation Set-up

All simulations have been implemented using MATLAB. Assuming that (between 100 nodes a 500) are randomly distributed in field, the base station is located at position (0, 0), provided with sufficient energy resources. Each node is equipped with an energy source which is set to 0.5J at the beginning of the simulation. We have set the percentage of CH between 5% and 25%. The simulation parameters are given in Table 1. The performance of the proposed O-LEACH protocol scheme is compared with that of Leach protocol.

- Orphan Nodes: percentage of connected orphan nodes.
- Network Life time: The number of nodes which are alive at the end of the simulation.
- Energy Consumption: the amount of consumed energy by the network in each round.

The base station is located at position (0, 0), provided with sufficient energy resources. Each node is equipped with an energy source which is set to 0.5J at the beginning of the simulation. We have set the percentage of CH between 5% and 25%.

| Parameter | Values |
|---|---|
| Simulation Round | 2000 |
| Sink Location | (0,0) |
| Network Size | 300 * 300 |
| Number of nodes | 500 |
| CH probability | 0,1 |
| Initial node power | 0.5 Joule |
| Nodes Distribution | Nodes are uniformly distributed |
| Data Packet size | 2000 bits |
| Energy dissipation ($E_{fs}$) | 10*0.000000000001 Joule |
| Energy for Transmission ($E_{TX}$) | 50*0.000000000001 Joule |
| Energy for Reception ($E_{RX}$) | 50*0.000000000001 Joule |
| Energy for Data Aggregation (EDA | 5*0.000000000001 Joule |

Table 1: Simulations parameters values.

## 4.2 Simulation Results

In this section, we present and discuss the simulation results of the protocol O-LEACH, to evaluate its performance and the execution of the protocol to show that the number of orphan nodes is almost null. Clustering algorithm distributed as O-LEACH requires that the sensor nodes are synchronized in their implementation. We carried out experiments on the percentage of the desired number of CH between 5% and 20%, the result we shows almost zero values. In the LEACH protocol, if the number of CH is very high, there will be a large number of nodes (CH) which consume a lot of energy. Thus, there will be considerable energy dissipation in the network. Also, if the number of CH is very small, the latter will manage groups of large sizes. Thus, this CH will be consumed rapidly in case an important work is required and there will be a risk of having a very large number of orphan nodes.

The necessary the timing to search for orphan nodes for a gateway. Each CH receives a message from a gateway and necessitates the timings for search orphan nodes.

The contribution of O-LEACH as compared to LEACH leads to a better network coverage of the whole controlled environment in several applications of WSNs. So, for each network organization (clusters during training set-up phase), we must calculate the total number of orphan nodes and the number of those who could reach a gateway. The ideal is that 100% of orphan nodes could be covered in the network. Moreover calculating the number of packets that arrive to the base station for both algorithms (O-LEACH and LEACH) shows that O-LEACH allows more data availability. O-LEACH is of a major importance, especially if the observed phenomenon occurs at orphan nodes.

The collected values by orphan nodes provide more availability of data at the base station. This enables a better decision in the application and a quick response. The simulation results show that our protocol outperforms the classic approach LEACH in terms of coverage.

In fig 7, show the distribution nodes in network Size 300*300 for number nodes=500. the orphaned nodes are concentrated in the end of the wireless sensor network. the number of gateway is less than the number of orphaned nodes. Each gatway can have one or more orphan nodes.

In fig 8, the number of orphan nodes is 30 in 100 rounds and the number of Gateway is 7 nodes. Coverage is complete; all orphaned nodes possess the gateway. O-LEACH provides total coverage of the work area.

In fig. 9, clearly shows the number of orphans is 60 knots in 400 rounds and the number of Gateway is 17nodes. Coverage is complete, all orphaned nodes possess the gateway. O-LEACH providers total coverage of the Works area.

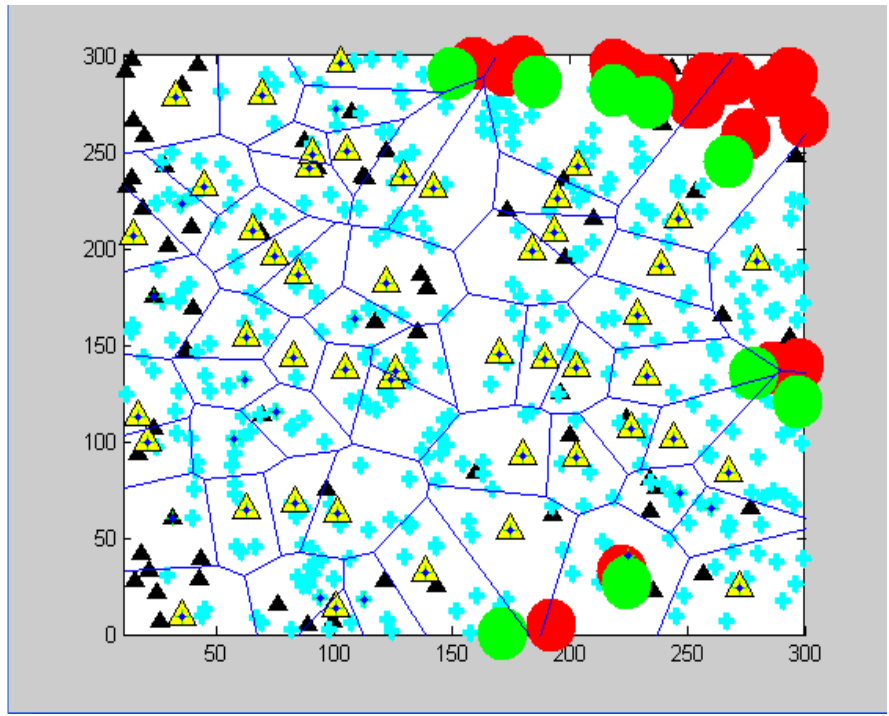

Figure 7. Distribution nodes: number of nodes 500 and Network Size 300*300

Clusyer Head

Dead Node

Node

Gateway

Orphan Node

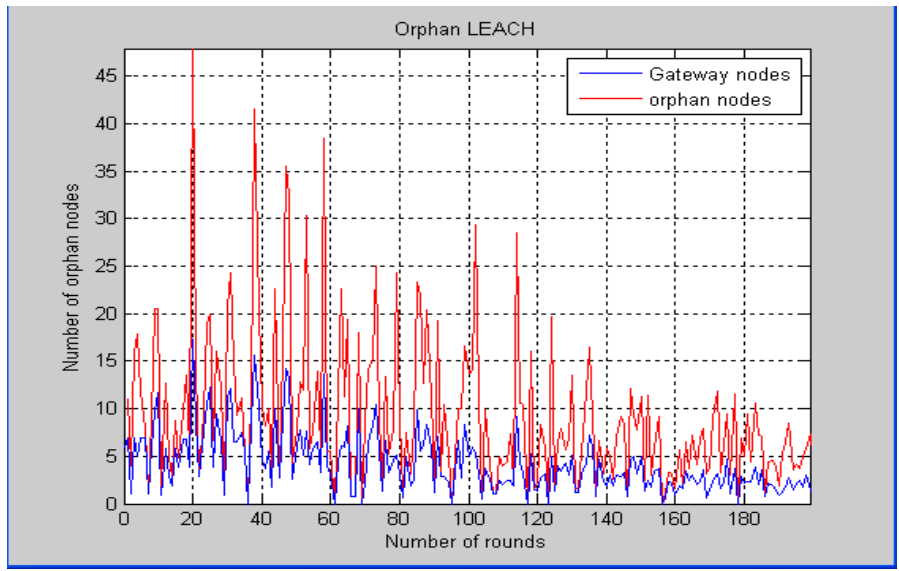

Figure 8. Number of Orphan nodes and gateway in 200 rounds

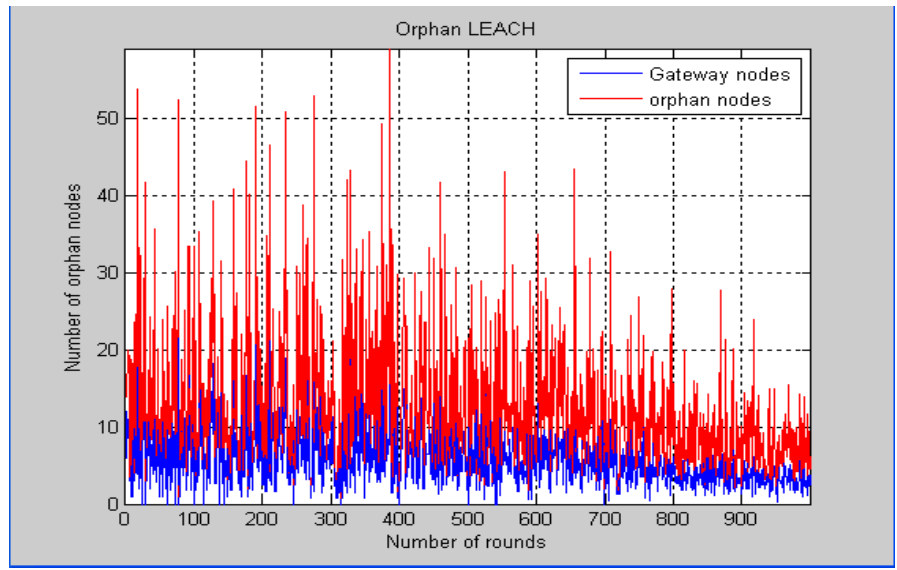

Figure 9.  Number of Orphan nodes and gateway in 1000 rounds

## 5. CONCLUSIONS

Several kinds of existing clustering protocols have been developed to balance and maximize the lifetime of the sensor nodes in wireless sensor networks. In this paper, we were interested in designing an O_LEACH routing protocol in order to minimize the orphan nodes in a round. It is estimated that the optimal and adequate percentage of the desired CH should be between 8% and 12% of the total number of nodes. Simulation results show that our protocol outperforms in terms coverage. Consequently, the cluster will be of a uniform size or each CH has a limited number of members. This involves optimizing the energy and the duration of lifetime of the networks. In this paper we explore the coverage of WSN nodes using Gateway to join between the orphan nodes and CH. for the selection of CH must have a minimum of energy, so that the CH allows managing the necessary communications with the member and base station. Through extensive simulations we obtained adequate and effective results, we have to reach 100% of total coverage. This protocol can improve the connectivity and of the network reliability with lower orphan nodes. The values collected by orphan nodes provide more availability of data at the base station. This enables a better decision on the application and a quick response.

## Authors

**Wassim Jerbi** is currently a Principal Professor in computer science at the Higher Institute of Technological Studies of Sfax –Tunisia. He is preparing his PhD in computer systems engineering at the National School of Engineers of Sfax –Tunisia where he is member of Computer and Embedded System Laboratory. His research and teaching interests focus on Wireless Sensor Networks, Routing Protocol.

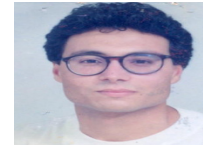

**Abderrahmen Guermazi** is currently a Technologist Professor in computer science at the Higher Institute of Technological Studies of Sfax –Tunisia. He is preparing his PhD in computer systems engineering at the National School of Engineers of Sfax –Tunisia where he is member of Computer and Embedded System Laboratory. He received the National Aggregation degree in computer science at 1998. His research and teaching interests focus on Wireless Sensor Networks, Routing and Security. He has several publications in international conferences of high quality.

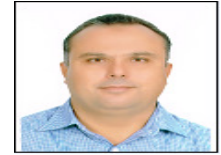

**Hafedh Trabelsi** received the B.S. degree from Sfax Engineering School (SES), university of Sfax, Tunisia, in 1989, the M.S. degree in the Central School of Lyon, France, in 1990, the Ph.D degree from the University of Paris XI Orsay, France, in 1994 and the habilitation degree from the University of Sfax, in 2008, all in Electrical Engineering, . He is currently a full professor of Electrical Engineering in (SES). He is a member of the research laboratory computer Embedded system (CES) dealing will smart system in device

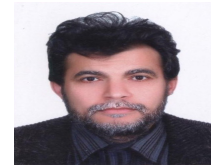